\documentclass{pasj01}
\Received{$\langle$reception date$\rangle$}
\Accepted{$\langle$acception date$\rangle$}
\Published{$\langle$publication date$\rangle$}

 \usepackage{natbib, aas_macros}
 \usepackage{bm}
 \usepackage{ulem}
\begin{document}
\title{
A radiation-hydrodynamic model of accretion columns for ultra-luminous X-ray pulsars}
 \author{Tomohisa \textsc{Kawashima} \altaffilmark{1,*},
 Shin \textsc{Mineshige} \altaffilmark{2},
  Ken \textsc{Ohs{uga} \altaffilmark{3,4},
  and Takumi \textsc{Ogawa} \altaffilmark{2}}}
\altaffiltext{1}{Center for Computational Astrophysics, National
 Astronomical Observatory of Japan, 2-21-1 Osawa, Mitaka, Tokyo 181-8588}
\altaffiltext{2}{Department of Astronomy, Graduate School of Science,
Kyoto University, Kitashirakawa, Oiwakecho, Sakyo-ku, Kyoto 606-8502}
 \altaffiltext{3}{Division of Theoretical Astronomy, National
 Astronomical Observatory of Japan, 2-21-1 Osawa, Mitaka, Tokyo 181-8588}
 \altaffiltext{4}{School of Physical Science, Graduate University of
 Advanced Study (SOKENDAI), Shonan Village, Hayama, Kanagawa 240-0193}
\email{kawashima@cfca.jp}

\maketitle

\begin{abstract}
Prompted by the recent discovery of pulsed emission from an
 ultra-luminous X-ray source, M82 X-2 ({\lq\lq}ULX-pulsar{\rq\rq}), we 
perform a two-dimensional radiation-hydrodynamic simulation of a super-critical 
accretion flow onto a neutron star through a narrow accretion column.
We set an accretion column with a cone shape filled with tenuous gas
with density of $10^{-4} {\rm g}~~ {\rm cm}^{-3}$ above a neutron star and
solve the two dimensional gas motion and radiative transfer within the column. 
The side boundaries are set such that
radiation can freely escape, while gas cannot.
Since the initial gas layer is not in a hydrostatic balance,
the column gas falls onto the neutron-star surface, thereby a shock being generated.
As a result, the accretion column is composed of two regions:
an upper, nearly free-fall region and a lower settling region, 
 as was noted  by Basko \& Sunyaev (1976).
The average accretion rate is very high; ${\dot M}\sim 10^{2-3} L_{\rm E}/c^2$ 
(with $L_{\rm E}$ being the Eddington luminosity), and so
radiation energy dominates over gas internal energy entirely within the column. 
Despite the high accretion rate, the radiation flux in the laboratory frame is 
kept barely below $L_{\rm E}/(4\pi r^2)$ at a distance $r$ in the settling region
so that matter can slowly accrete.
This adjustment is made possible, since large amount of photons produced via 
dissipation of kinetic energy of matter can escape through the side boundaries. 
The total luminosity can greatly exceed $L_{\rm E}$ by several orders of magnitude,
whereas the apparent luminosity observed from the top of the column is much less.
Due to such highly anisotropic radiation fields, 
observed flux should exhibit periodic variations with the rotation period, 
provided that the rotation and magnetic axes are misaligned. 
\end{abstract}

\section{Introduction}

The discovery of pulsed emission from the representative Ultra-luminous
X-ray sources (ULX), M82 X-2,
by \cite{2014Natur.514..202B} was really sensational, 
since it for the first time provided good piece of information regarding
the mass of the central objects in the ULXs. 
The ULXs are off-nuclear, compact, very luminous X-ray sources with X-ray
luminosities exceeding $\sim 10^{39}$ erg s$^{-1}$
\cite[see reviews
by][]{1989ARA&A..27...87F,2007Ap&SS.311..213S,2011NewAR..55..166F}.
There has been a long debate regarding the central engine of the ULXs:
it could be either sub-critical accretion onto an intermediate-mass black hole
\cite[see e.g.,][]{1999ApJ...519...89C,2000ApJ...535..632M}, or
super-critical accretion onto a normal stellar-mass black hole
\cite[see e.g.,][]{2001ApJ...552L.109K,2001ApJ...549L..77W,2006PASJ...58..915V},
or the combination of these two.
Although the direct mass measurement is the most effective way to settle down 
this issue, it is not always feasible due mainly to technical
difficulties.
In fact, ULXs are mostly very faint because of large distances.
That is, there had been no direct evidences in favor of or against 
the super-critical accretion scenario for ULXs until the discovery of
the ULX-pulsar. 
 Since it is obvious that the mass of a neutron star (NS) is less than $\sim$
 2 M$_\odot$, 
the discovery of the ULX-pulsar robustly points to the occurrence of
super-critical accretion, at least, in some ULXs. 
(We cannot ruled out a possibility that the ULXs may be a heterogeneous group.)

The discovery of the ULX-pulsar has re-opened the issue of super-critical 
accretion onto a magnetized NS. This is not totally a new issue but
was already intensively discussed in the 1970's, being
pioneered by \cite[][hereafter, BS76]{1976MNRAS.175..395B}. 
They solved one-dimensional structure of an accretion column, 
finding that the column is composed of two regions: an upper,
nearly free-fall region and a lower settling region. They also claimed
luminosities can naturally exceed the Eddington luminosity, since
copious photons produced via dissipation of kinetic energy of accreting
gas propagate towards side boundaries and can freely escape.
Large amounts of photons can thus leak away from the side of the column.
BS76 mainly addressed the super-critical accretion problem assuming hollow
columns, and the limiting luminosity sensitively depends on
the geometry of the 
accretion funnel, i.e., the thickness parameter of the wall of the column.
The two dimensional gas motion was not taken into account, 
and the super-critical column accretion without assuming the hollow
structure is not well understood.
In addition, they assumed a radiative diffusion approximation to
evaluate radiative flux at the side surface of the 
accretion column, which may 
underestimate the amount of the photons escaping from the side wall of
the column.

In this {\it letter}, we mainly address the following questions:
(1) How is super-critical accretion through a narrow accretion column on
a star feasible? (2) Can the luminosity of the ULX-pulsar be explained?      
(3) What observational signatures are expected for such a source? 
(4) Are there any substantial differences between the accretion processes 
onto black holes and those onto NSs?
In relation to these questions, it is of great importance to note that
\cite{2007PASJ...59.1033O}
solved the problem of super-critical accretion disk onto
non-magnetized NSs 
for the first time,  
by means of the two-dimensional radiation-hydrodynamic (RHD) simulations.
His main conclusions are that (1) super-critical accretion is indeed feasible and that 
(2) a large fraction of the liberated energy goes to kinetic energy
in NS accretion. By contrast,
more fraction of energy goes to radiation in black hole accretion.
In the case of accretion onto strongly magnetized NSs, however, flow
structure is distinct from that onto non-magnetized NSs: the strong
magnetic fields of the NSs penetrating the accretion disks can disrupt
the disk accretion and polar accretion flows
aligned with the B-fields of the NSs can be formed inside 
Alfv${\acute {\rm e}}$n radii.

To examine whether the situation changes or not in the case of column
accretion, we solve the gas dynamics and radiation transfer inside the funnel
by two-dimensional RHD simulations. 
We expect much larger radiation output from the side wall
than that from the top of accretion column, thus very high
accretion luminosities being expected. 
Such geometrical differences will be essential from the viewpoints of radiation
hydrodynamics, as will be demonstrated in this
{\it letter}.  

\section{Our Model and Numerical Methods}

We apply
the simulation by \cite{2007PASJ...59.1033O} to the case of NS accretion
through a narrow accretion column (or funnel).
We set a calculation box with a cone shape having a half opening 
 angle of $\pi/6$ on the NS surface and 
 fill it with uniform tenuous gas with density of
 $\rho=10^{-4} {\rm g} ~{\rm cm}^{-3}$
 and high temperature of $T_{\rm gas}=10^7$ K.
We further assume that the initial radiation temperature surrounding the column
is low, $T_{\rm rad} = 10^4$ K, and that gas has no angular momentum,
for simplicity. 
The mass of a NS is set to be $M=1.4 M_\odot$.

Since the gas layer within the column is not in hydrostatic balance, 
gas starts to fall onto the lower boundary after the start of a simulation.
We calculate two dimensional gas motion and radiation transfer by
solving a set of RHD equations in the spherical coordinates 
 ($r$, $\theta$, $\varphi$), assuming axisymmetry.
The basic equations are
 \begin{equation}
  \frac{{\partial}\rho}{{\partial}t} +
   {\nabla}{\cdot}\left({\rho}{\bm v}\right) = 0,
 \end{equation}
  \begin{equation}
   \frac{{\partial}({\rho}v_{r})}{{\partial}t} +
    {\nabla}{\cdot}\left({\rho}v_{r}{\bm v}\right) =
    -\frac{{\partial}p}{{\partial}r} +
    {\rho}\left[\frac{v_{\theta}^2}{r} -
   \frac{GM}{(r - r_{\rm s})^2}\right] + \frac{\chi}{c}F^{r}_{0},
  \end{equation}
  \begin{equation}
   \frac{{\partial}({\rho}rv_{\theta})}{{\partial}t} +
    {\nabla}{\cdot}\left({\rho}rv_{\theta}{\bm v}\right) =
    -\frac{{\partial}p}{{\partial}{\theta}} + r\frac{\chi}{c}F^{\theta}_{0},
  \end{equation}
  \begin{equation}
   \frac{{\partial}e}{{\partial}t}+{\nabla}{\cdot}(e{\mbox{\boldmath$v$}})
    =-p{\nabla}{\cdot}{\mbox{\boldmath$v$}}-4{\pi}{\kappa}B+c{\kappa}E_{0},
  \end{equation}
  \begin{equation}
   \frac{{\partial}E_{0}}{{\partial}t}+{\nabla}{\cdot}(E_{0}{\mbox{\boldmath$v$}})
    =-{\nabla}{\cdot}{\mbox{\boldmath$F$}_{0}}
    -{\nabla}{\mbox{\boldmath$v$}}{\colon}{\mathbf{P}}_{0}
    +4{\pi}{\kappa}B-c{\kappa}E_{0}.
  \end{equation}
Here, $\rho$ is the mass density, ${\bm v}
=(v_{r},~v_{\theta},~v_{\varphi})$ is the velocity
(we set $v_{\varphi} =0$, assuming no angular momentum of gas), 
$p$ is the gas pressure, $e$ is the internal energy
density of gas, $B$ is the blackbody intensity, $E_{0}$ is the radiation
energy density of gas, ${\bm
F}_{0}=(F^{r}_{0},F^{\theta}_{0})$ is the radiative flux, ${\bf P}_{0}$ is
the radiation pressure tensor (all of the quantities with suffix $0$ are
measured in the comoving frame of the fluid), ${\kappa}$ is the absorption opacity,
and ${\chi} = {\kappa} + {\rho}{\sigma}_{\rm T}/m_{\rm p}$ is the total
opacity, where ${\sigma}_{\rm T}$ is the cross section of Thomson
scattering and $m_{\rm p}$ is the proton mass.
We do not take into account thermal Compton
cooling/heating, which, if included, might change the
temperature structure near the NS surface, though the overall
structure should not alter significantly because of a very large optical
depth of the column.
General relativistic effects are incorporated by a pseudo-Newtonian
potential prescribed by \cite{1980A&A....88...23P},
${\Psi}=-GM/(r-r_{\rm s})$.  
 Here, $r_{\rm s}(=2GM/c^2)$ is the Schwarzschild radius, 
 $G$ is the gravitational constant, and $c$ is the speed of light.
 Simulations are carried out using a RHD code
 \citep{2005ApJ...628..368O,2007PASJ...59.1033O},
  which solves the radiative transfer by adopting the flux-limited
diffusion (FLD) approximation
\citep{1981ApJ...248..321L,2001ApJS..135...95T} and hydrodynamic
equations by using Virginia Hydrodynamics One (VH-1) based on
Lagrange-remap version of the piecewise parabolic method 
\citep[PPM,][]{1984JCoPh..54..174C}, which is the third order Godunov's scheme.

The size of the computational domain is $r_{\rm in}$ ${\le}$ $r$ ${\le}$
$r_{\rm out}$  and $\theta_{\rm in}$
  ${\le}$ ${\theta}$ ${\le}$ $\theta_{\rm out}$, where
$(r_{\rm in}, ~ r_{\rm out}) = (2.4r_{\rm s}, ~ 500r_{\rm s})$ and
$(\theta_{\rm in}, ~ \theta_{\rm out}) = (0, ~ {\pi}/6)$; that is, we
set $r_{\rm in} \sim 10$ km, typical NS radius.
We set a rather long accretion column (with a height of 500 $r_{\rm s}$)
so as to include the trapping radius, which is expected to be at
$r~\sim~ (L/L_{\rm E})r_{\rm s}$ \citep{1978MNRAS.184...53B}, within the
calculation box.
(We will show later $L/L_{\rm E} ~ \sim 10^2$.)
However, we focus our discussion in a narrow region, within 40 km
($\sim 10 r_{\rm s}$).

The number of grid cells is
$N_{r} {\times} N_{\theta} =288 {\times} 96$.
 The grid points are distributed such that
${\Delta}{\log}~r = ({\log}~r_{\rm out} - {\log}~r_{\rm in})/N_r$ 
and
${\Delta}{\theta} =1/N_{\theta}$.

At $r=r_{\rm in}$ the accreting gas is absorbed, however, 
we assume that the energy carried by the gas is  
immediately converted to the radiation
and is released as outgoing radiation flux \citep{2007PASJ...59.1033O}.
The top of the computation box at $r=r_{\rm out}$ is set to be a free boundary, 
where both inflow and outflow of gas is allowed, i.e., the fluid
quantities with no gradient are assigned to the ghost cells at every time step.
We impose the mirror boundary condition both at $\theta = \theta_{\rm in}$
and $\theta_{\rm out}$ for the gas.
As for the radiation, we set the mirror boundary at $\theta = \theta_{\rm in}$,
while we let radiation go out through the boundary at $\theta = \theta_{\rm out}$
with radiation flux, $F^{\theta}_{0} = cE_{0}$.

We set the initial mass density to be $\rho = 10^{-4} {\rm g}
 ~{\rm cm}^{-3}$ so as to 
 reproduce the luminosity of M82 X-2, $L \sim 10^2 L_{\rm E}$ for
 neutron stars with $1.4M_{\odot}$, i.e.,
 ${\dot M} \sim 10^{20} {\rm g} ~ {\rm s}^{-1}$. Here, we have assumed
 the radiative efficiency $\sim 0.1$.
 By using 
 the relation ${\dot M} \simeq 4 \pi r^2 {\rho} v_{\rm ff}
 (\Omega/4\pi)$, we can obtain ${\rho} \sim 10^{-4}$ ${\rm g}
~ {\rm cm}^{-3}$ for a radius inside which matter can accrete enough by
 the end of our simulation
 $r \sim 5 \times 10^7 {\rm cm}$, free-fall velocity there $v_{\rm ff} \sim 3 \times 10^8 {\rm cm} ~{\rm s}^{-1}$, and
 the solid angle subtended by the accretion 
  column ${\Omega} {\sim} 0.1 \times (4\pi)$.
 In fact, we confirm that the simulated
 luminosity is ${\sim} 10^{40} {\rm erg} ~{\rm s}^{-1}$, which is
consistent with the estimated luminosity of M82 X-2.

\section{Results}
As was mentioned before, gas starts to freely fall towards the
NS immediately after a simulation starts, since
there is no sufficient pressure force preventing the gas fall 
in the initial state.
The mass accretion rate onto the NS shows $\sim 10^{2-3} L_{\rm E}/c^2$, as
expected above.

  \begin{figure*} [!th]
  \begin{center}
      \includegraphics[scale=0.59]{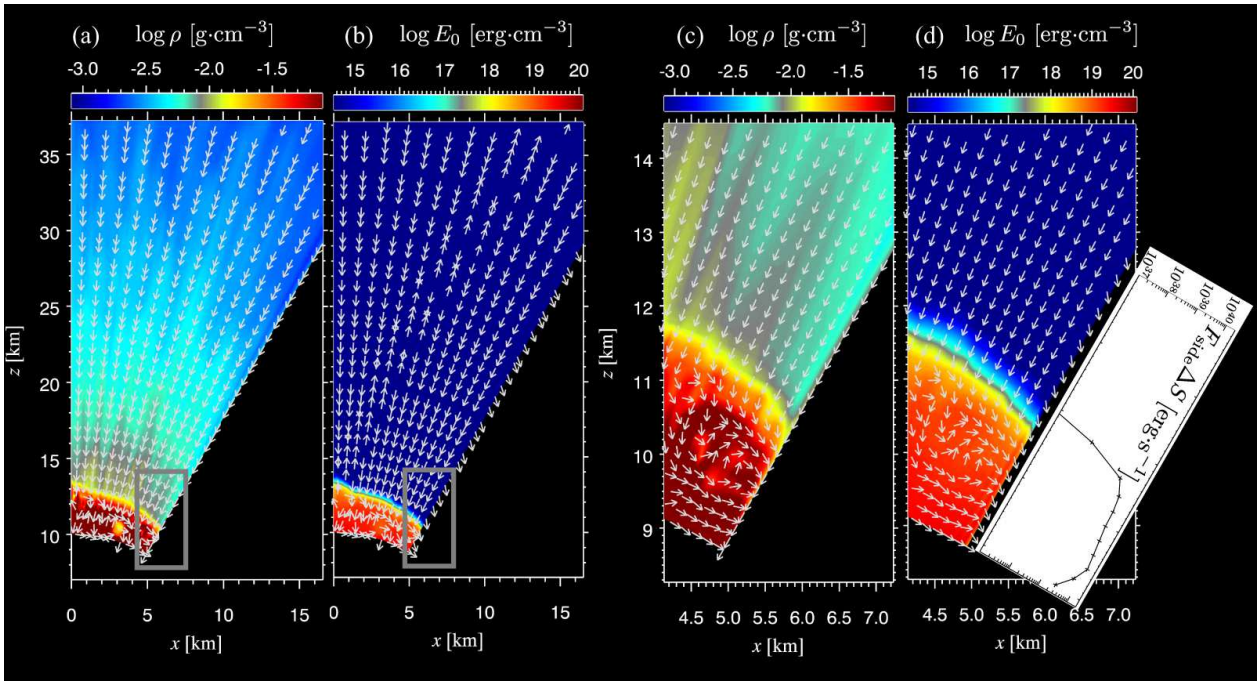}
  \end{center}
     \caption{Two-dimensional diagrams displaying mass and energy flow
   of super-critical  
      column accretion at the elapsed time of 0.0335 sec.
   The left two panels show the structure of a  
   column within $r=37 {\rm km}$, whereas the right two are magnified views
   of the innermost region enclosed by the gray squares in the left ones. In
   each pair of panels, the left panels show 
   matter density color contours overlaid with matter velocity, while the
   right ones show color contours of 
   radiation energy density overlaid with radiation flux in the laboratory frame
   (i.e., ${\bm F_0} + {\bm v} E_0$), respectively.
   Radial profile of the radiation luminosity leaked from the side of
   accreting column with an area of $\Delta S \equiv 2\pi r \sin
   \theta \Delta r$ is 
   shown in the inserted figure, where $\Delta r~ \sim ~ 0.2$ km is 
   the mesh spacing.}
   \label{fig1}
  \end{figure*}

  We first show snapshots displaying gas and radiation properties
over the two-dimensional plane in figure \ref{fig1}.
These are color contours of matter density overlaid with gas velocity [panels (a) and (c)]
 and those of radiation energy density overlaid with comoving radiation flux
 [panels (b) and (d)], respectively. The right two panels are magnified views
of the innermost parts
of the corresponding left panels.

Two-zone structure is very clear in these plots. The interface between the
two zones is located at around $r=13 ~{\rm km} ~\equiv~ r_{\rm shock}$,
where a shock structure is observed there; 
velocity vectors are uniformly downward in the upper free-fall region,
whereas they show significant deviations from the uniform down-flow
below the interface.
Even circular motion is observed there.

Another important feature found in this figure is that most of radiation
originates from the side wall of the lower settling region of the accretion
column, not from the top of the column.
Interestingly, the luminosity of the NS surface is kept barely
below $L_{\rm E}\times\Omega/(4\pi)$ so that mass accretion onto 
the NS can proceed, liberating accretion energy sideway.
The inset figure in panel (d) of figure \ref{fig1} clearly shows that
the radiation flux profile $ F_{\rm side} \equiv
F^{\theta}_0(r,\theta_{\rm out})$ is nearly flat along the side wall below the
interface.
We see that the fractional luminosity $F_{\rm side} \Delta S$ 
exceeds the Eddington luminosity ($\simeq 2 \times 10^{38}$ erg
s$^{-1}$) at $r < r_{\rm shock}$,
where $\Delta S \equiv 2 \pi r \sin \theta \Delta r$ with $\Delta r$
being the mesh  
spacing ($\Delta r \approx 0.2$ km near the NS surface).
The total luminosity amounts to $L\sim 6 \times 10^{40}$ erg ~ s$^{-1}$
$\sim 300 L_{\rm E}$, as will be explicitly shown later.

\begin{figure} [!t]
  \begin{center}
         \includegraphics[scale=0.35]{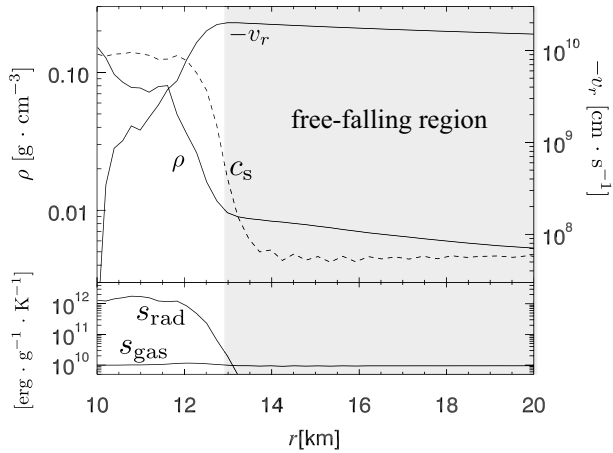}
  \end{center}
    \caption{(Top) Azimuthally averaged radial profile of the matter
    density and the velocity at the elapsed time of $t=0.0335$ sec. 
    Each quantity is averaged in the
    $\theta$ direction, e.g., $2\pi \int^{\theta_{\rm out}}_{0} \rho r^2
    \sin{\theta} d{\theta}/ (2\pi \int^{\theta_{\rm out}}_{0} r^2
    \sin{\theta} d{\theta})$. 
    A weak shock formed at $r~{\sim}~13$ km separates the upper, nearly
    free-fall region (shaded region) and the settling region, in which
    gas is 
    drastically decelerated by the radiation 
    pressure and matter is thus being accumulated.
    (Bottom) Same as the top panel but for gas entropy and radiation entropy.
    }\label{fig2}
   \end{figure}

In order to check how physical quantities vary as matter falls in the
settling region,
we plot the azimuthally averaged gas density and velocity in figure
 \ref{fig2}. 
We find that mater density gradually increases downward 
by about two orders of magnitudes, while
the absolute value of gas velocity decreases downward by roughly the
same orders of magnitudes.
These gradual changes arise, since we azimuthally averaged the physical
quantities, and since the shock surface is not parallel to
NS surface but is tilted (see figure \ref{fig1}). 
Note also that the radiation entropy displayed in the bottom panel shows a
hump with a plateau shape below the shock region,
indicating that energy dissipation really occurs there and that
dissipated energy directly goes to radiation with going very little to
gas. (Note that the gas entropy is very small, compared with the
radiation entropy.)
The plateaus shape indicates that the radiation gains entropy from gas
and loses entropy by outgoing radiation at similar rates.

\begin{figure} [!t]
  \begin{center}
      \includegraphics[scale=0.4]{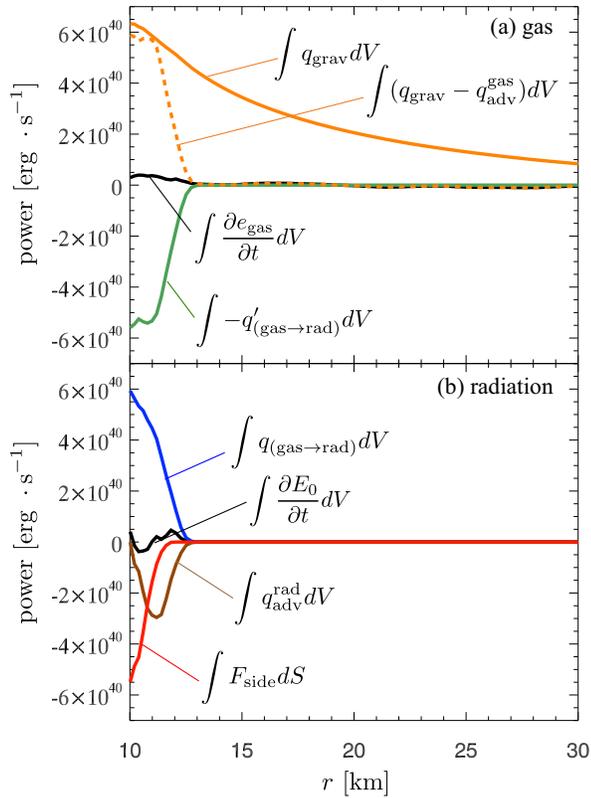}
  \end{center}
    \caption{ Radial distributions of energy density forms of gas 
    (upper panel) and of radiation (lower panel), respectively.
    The elapsed time is $t=0.0335$ sec.
    }\label{fig3}
   \end{figure}

   We have so far confirmed that super-critical
   accretion actually 
   occurs through the accretion column. The next issue is to clarify how it
   occurs despite  
   huge radiation energy density accumulating on the NS surface.
   For this purpose, we examine how the energy conversion occurs from
   gas to radiation by using figure \ref{fig3}.
   Here, the plotted quantities are either of volume-integrated energy
   density or surface-integrated energy flux density defined as:
$ \int q dV \equiv
  2\pi \int^{\bar r}_r \int^{\theta_{\rm out}}_{\theta_{\rm in}}q ~r^2 \sin \theta
  dr d\theta$,
and
$ \int F_q dS \equiv
  2\pi \int^{\bar r}_r F_q \sin \theta_{\rm out} ~r dr$,
where $q$ is energy density, $F_q$ is energy flux density and
integration is made from radius $r$ to ${\bar r} = 50$ km.
The selected quantities are gravitational energy ($q_{\rm grav}$), 
 advection energy of gas ($q_{\rm adv}^{\rm gas}$),
 time derivative of gas energy density ($\partial e_{\rm gas}/\partial t$), 
 radiation energy density converted from gas
 (with two different expressions, $q^{\prime}_{({\rm gas} \rightarrow {\rm rad})}$
  and $q_{({\rm gas} \rightarrow {\rm rad})}$),
time derivative of radiation energy density ($\partial E_0/\partial t$),  
advection energy of radiation ($q^{\rm rad}_{\rm adv}$), 
 and radiation energy flux density emitted from the unit area of the side wall
  ($F_{\rm side}$), respectively, where
$q_{\rm grav} \equiv -GM v_r/(r - r_{\rm s})^2$, 
$q_{\rm adv}^{\rm gas} \equiv 
        {\bm \nabla} \cdot (\rho v^2 {\bm v}/2 + \gamma e {\bm v})$, 
$e_{\rm gas} \equiv \rho v^2/2 + e$, 
$q^{\prime}_{({\rm gas} \rightarrow {\rm rad})} \equiv
     4 \pi \kappa B - c \kappa E_0 + (\chi / c){\bm F}_0 \cdot {\bm v}$,
$q_{({\rm gas} \rightarrow {\rm rad})} \equiv
     4 \pi \kappa B - c \kappa E_0 + {\bm \nabla}{\bm v}:{\bf P}_0$, 
and $q^{\rm rad}_{\rm adv} \equiv {\bm \nabla} \cdot ({\bm v}E_0)$, 
respectively.

Let us first see how gas energy changes its form in the upper panel of
figure \ref{fig3}. 
As gas falls down, the gravitational energy is gradually released
(see the line labeled with $\int q_{\rm grav}dV$).
However, the sum of the liberated energy and the advection energy is
kept nearly constant until gas reaches the shock surface at 
$r_{\rm shock} \sim 13$ km, meaning that all the liberated energy is carried
downward by the falling gas.
Since photons near the NS push the plasma outward, 
falling gas is suddenly decelerated at the shock surface.
Below the shock surface, a part of the liberated energy is converted to
radiation energy (see the line labeled with
   $\int - q^{\prime}_{({\rm gas} \rightarrow {\rm rad})}dV$).
As a result, the gas energy $e_{\rm gas}$ is kept nearly constant in time,
which guarantees the existence of stationary flow.

Let us next see how radiation energy changes its form in the lower panel
of figure \ref{fig3}. 
As mentioned above, gas energy is converted to radiation energy below
$r_{\rm shock}$  (see the line labeled with
$\int  q_{({\rm gas} \rightarrow {\rm rad})}dV$). 
Nearly the same amount of energy is radiated away through the side boundary
(see the line labeled with $\int F_{\rm side} dS$).
That is, the converted energy to radiation is carried to the side boundary
and is radiated away at a similar height.
We note, however, that the time derivative of the radiation energy density
($\partial E_0/\partial t$) is non-zero below $r_{\rm shock}$ and
is nearly balanced with the advection of radiation energy. 
This occurs because of convective motion of radiation bubbles
[see circulating  vectors in fig. 1 (c) and (d)].

The matter under the shock slowly accretes onto the NS, and emit the
photons, whose luminosity is nearly Eddington luminosity.
We note that isotropic luminosity at the NS surface is
${\sim}3L_{\rm E}$, which can be obtained by multiplying
the luminosity at the NS surface within the accretion column
$2\pi\int_0^{\theta_{\rm out}}F_0^r(r_{\rm 
in},\theta)r^2 \sin{\theta}~d{\theta}$ by the correction factor
 $4\pi/{\Omega}$.
This isotropic luminosity is consistent with that of
\cite{2007PASJ...59.1033O}.

\section{Discussion}

\begin{figure} [!t]
  \begin{center}
   \includegraphics[scale=0.45]{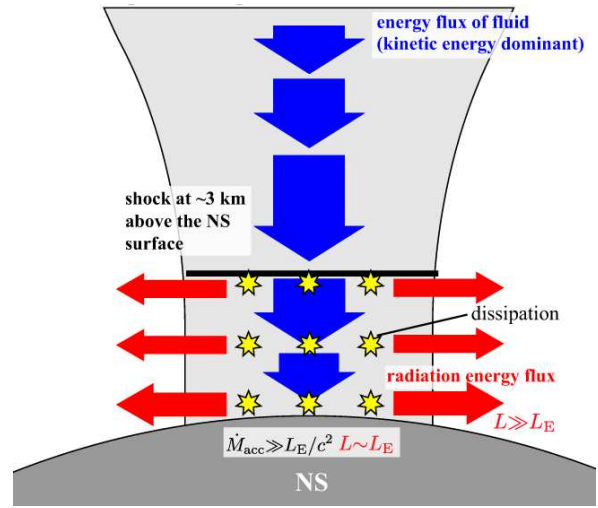}   
  \end{center}
    \caption{Schematic picture explaining energy flow from gas (potential energy)
    to outgoing radiation within a super-critical accretion column.
    The blue arrows represent the energy flow carried by gas; their
    length and width are drawn in proportional to the kinetic energy flux
    and the mass accretion rates, respectively, 
    whereas the red arrows represent energy flow carried by radiation and their 
    widths are drawn in proportional to the radiation energy flux. 
    }\label{fig4}
 \end{figure}

 
 In this $letter$ we study the two dimensional structure of a narrow
super-critical accretion column with radiation transfer by means of
the axisymmetric radiation hydrodynamic simulation.
The accretion column is composed of an upper, free-fall region
and a lower settling region with a shock being formed at the interface.
Gas energy is converted to radiation only below the shock surface, and
it is finally radiated away from the side surface.
Super-Eddington luminosity is produced there, while
nearly Eddington luminosity is emitted from the NS surface.
Such basic energy flow is summarized in figure \ref{fig4}.

There are similarities and differences among three super-critical
accretion systems (see table 1).
The most distinctive features are found in the final form of energy;
that is, it differs among them how excess energy can be taken away
from the central engine at a rate exceeding the classical Eddington limit.
Excess energy is carried by trapped photons in the black hole disk accretion,
by outflow material in the form of its kinetic energy
in disk accretion onto a non-magnetized NS
\citep{2007PASJ...59.1033O}, and by outgoing radiation  
emitted from the side boundary in column accretion onto a magnetized NS.
In any case the outward flux measured in the laboratory frame 
(${F_0} + {v} E_0$) is much reduced in the innermost part of
accretion flow.
This condition makes super-critical accretion feasible.

  \begin{table}[b]
    \tbl{Three classes of super-critical accretion systems}{
   \begin{tabular}{lll}
    \hline
    central object & photon luminosity & final energy carrier\\
    \hline
    \hline
    black hole & $ \gtrsim L_{\rm E}$ & trapped photons \\
    non-magnetized NS  & $\sim L_{\rm E}$ & outflow (kinetic energy) \\
    magnetized NS & $ \gg L_{\rm E}(\Omega/4\pi)$ & outgoing radiation \\
    \hline
   \end{tabular}}
  \end{table}

Since most of excess energy is finally radiated away from the side boundary
in the magnetized NS case, strong electromagnetic radiation 
should be finally observed, unlike other two cases.
The most unique feature of the ULX-pulsar is highly anisotropic
radiation fields.
This occurs, since large luminosity radiation goes out through the side boundary
and not from the top of the accretion column. 
If the rotation axis and the magnetic axis of the NS do not coincide,
we thus expect pulsed emission to be observed.

In the present study, we observe circulating gas motion in the settling region
(see figure \ref{fig1}).
Such motion affects energy transport processes and is
described by the photon advection term in our analysis (see the bottom
panel of figure \ref{fig3}).
Such gas motion, known as {\lq\lq}photon bubbles{\rq\rq}, was
intensively studied  by \cite{1989ESASP.296...89K} by means of the RHD
simulations.
They carefully solve the super-Eddington atmosphere on the magnetized NS
and found strong radiation-driven, optically thin outflow
embedded in optically thick, inflowing plasma.
They estimated that the growth timescale of such photon bubbles are
is on the order of a millisecond. They also observe rapid time
variability on millisecond timescales.
\cite{1992ApJ...388..561A} made the linear analysis, finding that
the growth time is inversely proportional to the wavenumber,
thereby concluding that
the layer is eventually dominated by a few large bubbles.
These features are in good agreement with our results.

Despite the existence of internal gas circulation, the fractional
luminosity $F_{\rm side} {\Delta}S$
is remarkably uniform except in the innermost region and thus seems
not to be affected by photon bubbles [see the inserted figure next to
figure 1(d)].
The photon bubbles are time-dependent, thus photon advection directly leads
to time changes of radiation energy density (see the bottom panel of figure \ref{fig3}).
Such time dependence is smoothed on long terms and, hence,
does not affect the fractional luminosity profile.

In the present study, 
we assume that magnetic fields are not so strong so that the
two-dimensional gas motion can be allowed 
within the column.
That is, in our model, 
both of radiation-pressure and ram-pressure forces
are comparable to the Lorentz force and transport the radiative energy
by the advection, while the accretion column is possibly sustained due
to marginally strong Lorentz force at the side surface of the column.
By using our simulation data, this marginal B-field
strength inside the column can be estimated to be $\sim 10^{10}$G, which
is not anomalous in pulsars. 
 The perfect confinement of the column may look
 slightly difficult to be sustained, since the radiative force can
 dominate the Lorentz force with the marginal magnetic fields at the
 side wall. However, the column accretions as shown in our simulation
 should be still feasible for the following two reasons: (1) The
 magnetic fields inside the column will be pushed towards the side wall
 due to the fluid motion with the frozen-in magnetic fields, leading to
 accumulation of the B-field at the column side. Then the enhanced
 magnetic tension force and magnetic pressure gradient force  can
 overcome the radiative force. (2) Even if the perfect confinement of
 the column is not sustained, the mass loss due to radiative force in
 the $\theta$-direction at the side wall cannot significantly change our
 result, since the sound speed of the radiation dominated gas at the
 side surface is ${\sim}10$\% of the accretion velocity, indicating that the
 mass loss rate through the column side will only be $\sim 10$\%, at maximum,
 of mass accretion rate.
 We plan to explore as future work the cases of much stronger magnetic
 fields which forbid the fluid motions perpendicular to the magnetic
 field (i.e., the $\theta$-direction).

Here, we discuss the location of the photosphere. In the
$r$-direction, the scattering photosphere and the effective photosphere
appear at $\sim  0.03$ km and $\sim 100$ km, respectively, inside from the
outer boundary ($r_{\rm out} \sim 2000$ km). Both of them are thus
located far outside 
the shock surface ($r_{\rm shock} \sim 13$ km). 
In the $\theta$-direction, the scattering photosphere is at
around $\sim 10^{-4}$ km from the side boundary at
$r \sim 12 {\rm km} (< r_{\rm shock})$.
Although the column is effectively optically thin
($\tau_{\rm eff} \sim 0.03$) at $r \sim 12$ km if the gas temperature is
kept high, the effective optical depth is more likely to be 
$\tau_{\rm eff} \simeq 10$ if
Comptonization is properly taken into account; that is, the matter
temperature will then be much cooler $\sim 10^{8}$ K (i.e., gas
temperature will decrease to radiation temperature via the Comptonization).
That is, the effective photosphere will appear around 0.5 km from
the side wall, if the gas is strongly Comptonized.
The Comptonization timescale is, in fact, much shorter than the accretion
timescale as briefly discussed below.


In the present study, we rather simplified the physical situation to focus
on the basic observational features of the
super-critical accretion column, 
thus omitting some potentially important physical processes.
For example, we ignored thermal Compton scattering,
which could be of great importance when considering detailed radiation spectra
\citep{2009PASJ...61..769K,2012ApJ...752...18K} because the plasma
is scattering dominant and gas temperature is much greater than
radiation temperature under the shock surface. (We note that the fluid dynamics found in this {\it letter} will not be
significantly affected by the Comptonization as described below.)
At first, we show the importance of the Comptonization to determine
gas temperature distribution, and then, we present the reason why the
dynamics will not be significantly modified by the Comptonization.
Under the shock surface, gas temperature is orders of
magnitude higher than radiation temperature.
The Comptonization, which relaxes the gas and radiation
temperature, will sufficiently change the gas temperature distribution,
because the estimated Comptonization timescale  ($\sim 10^{-11}$s) is by
seven orders of magnitude shorter than the accretion timescale($\sim 10^{-4}$
s). Here, the Comptonization timescale is estimated by using the Compton
cooling rate based on the Kompaneets equation
\citep[see, e.g. the equation
(3) in][]{2009PASJ...61..769K},
substituting gas temperature $\sim 10^{10}$K, radiation temperature
$\sim 10^{8}$K, and mass density $\sim 10^{-1.5} {\rm g}~{\rm cm}^{-3}$. 
Thus, we can expect that the gas temperature will significantly decrease
due to the Comptonization.
Despite the importance of the Compton scattering, we can expect that the
dynamics of the flow will not be significantly changed by the Comptonization.
This is because the flow dynamics is governed by the
radiative force and the radiation energy density is 1-3 orders
of magnitude larger than the gas internal energy density below the shock
surface, which indicates that the increment of the radiation energy
converted from the gas internal energy via the Comptonization is too
small to significantly change the dynamics of the accretion column.
The Comptonization may be, therefore, important only when we calculate
the spectra of the accreting column:
we expect that a modified-blackbody spectrum with a Wien peak will be formed,
since the accretion column is scattering dominant and the Compton {\it
y}-parameter is much greater than unity under the shock surface.
In this {\it letter},
three dimensional gas motion (due to the NS rotation, etc.) is
also ignored.
Further, we used the FLD approximation for solving radiation transfer,
which need to be improved.
We also simplify the plasma processes near the NS; that is,
we have not considered neutrino emission nor included corrections to
hydrodynamics and radiative transport 
due to strong magnetic fields \citep{1989ESASP.296...89K}.
These effects will be explored in a next paper.

  \begin{ack}
   We thank H.R. Takahashi for useful discussion.
The numerical simulations were carried out on the XC30 at the Center for
  Computational Astrophysics, National Astronomical Observatory of
  Japan.
This work is supported in part by Grants-in-Aid of the Ministry
of Education, Culture, Sports, Science and Technology (MEXT) (26400229,
   S.M.; 15K05036, K.O.).
 This work was supported in part by MEXT HPCI
STRATEGIC PROGRAM and the Center for the Promotion
   of Integrated Sciences (CPIS) of Sokendai, 
   and MEXT as a priority issue
   (Elucidation of the fundamental laws and evolution of the universe)
   to be tackled by using post-K Computer and JICFuS.  
  \end{ack}


\end{document}